\documentclass[journal=jpclat,amsmath,amssymb,manuscript=article]{achemso}

\usepackage{graphicx}
\usepackage{float}
\usepackage{dcolumn}
\usepackage{color}
\usepackage{latexsym,bm}
\usepackage[normalem]{ulem}
\usepackage{multirow}
\usepackage{appendix}

\usepackage{amsmath}

\author{Yuyang Ji}
\affiliation{Key Laboratory of Quantum Information, University of Science and
  Technology of China, Hefei, Anhui, 230026, People's Republic of China}
\affiliation{Synergetic Innovation Center of Quantum Information and Quantum
  Physics, University of Science and Technology of China, Hefei, 230026, People's Republic of China}
\author{Peize Lin}
\email{linpeize@sslab.org.cn}
\affiliation{Songshan Lake Materials Laboratory, Dongguan 523808, Guangdong, China}
\affiliation{Institute of Physics, Chinese Academy of Sciences, Beijing 100190, China}
\author{Xinguo Ren}
\email{renxg@iphy.ac.cn}
\affiliation{Songshan Lake Materials Laboratory, Dongguan 523808, Guangdong, China}
\affiliation{Institute of Physics, Chinese Academy of Sciences, Beijing 100190, China}
\author{Lixin He}
\email{helx@ustc.edu.cn}
\affiliation{Key Laboratory of Quantum Information, University of Science and
  Technology of China, Hefei, Anhui, 230026,  People's Republic of China}
\affiliation{Synergetic Innovation Center of Quantum Information and Quantum
  Physics, University of Science and Technology of China, Hefei, 230026, People's Republic of China}

\title[]{Importance of exact exchange to the geometric and electronic structures of   Cs$_2$$B$$B'$$X_6$  double perovskites}

\begin{document}

%\title{First-principles study of the  Cs$_2$$B$$B'$$X_6$  double perovskites}
%\title{Importance of exact exchange to the geometric and electronic structures of   Cs$_2$$B$$B'$$X_6$  double perovskites}
%\author{Yuyang Ji}
%\affiliation{Key Laboratory of Quantum Information, University of Science and
%  Technology of China, Hefei, Anhui, 230026, People's Republic of China}
%\affiliation{Synergetic Innovation Center of Quantum Information and Quantum
%  Physics, University of Science and Technology of China, Hefei, 230026, People's Republic of China}
%\author{Peize Lin}
%\email{linpeize@sslab.org.cn}
%\affiliation{Songshan Lake Materials Laboratory, Dongguan 523808, Guangdong, China}
%\affiliation{Institute of Physics, Chinese Academy of Sciences, Beijing 100190, China}
%\author{Xinguo Ren}
%\email{renxg@iphy.ac.cn}
%\affiliation{Songshan Lake Materials Laboratory, Dongguan 523808, Guangdong, China}
%\affiliation{Institute of Physics, Chinese Academy of Sciences, Beijing 100190, China}
%\author{Lixin He}
%\email{helx@ustc.edu.cn}
%\affiliation{Key Laboratory of Quantum Information, University of Science and
%  Technology of China, Hefei, Anhui, 230026,  People's Republic of China}
%\affiliation{Synergetic Innovation Center of Quantum Information and Quantum
%  Physics, University of Science and Technology of China, Hefei, 230026, People's Republic of China}

\begin{abstract}
We investigate the lead-free halide double perovskites (HDPs) Cs$ _2BB'X_6$ ($B$=Ag, Na; $B'$=In, Bi; $X$=Cl, Br) via first-principles calculations. We find that both the geometric and electric structures of the HDPs obtained by the  Heyd-Scuseria-Ernzerhof (HSE) hybrid functional are much better than those of the Perdew-Burke-Ernzerhof (PBE) functional. Importantly, we find that the electronic structures of DHPs are very sensitive to their geometries, especially the $B$-$X$ bond lengths. As a consequence, the electronic structures calculated by the HSE functional using the PBE optimized geometries may still significantly underestimate the band gaps, whereas the calculations on the HSE optimized geometries provide much more satisfactory results. The sensitivity of the band gaps of the DHPs to their geometries opens a promising path for the band structure engineering via doping and alloying. This work therefore provides an useful guideline for further improvement of HDPs materials.
\end{abstract}
\maketitle

%\section{Introduction}

Hybrid perovskites (HPs) exhibit excellent transport and optical properties, such as high carrier mobility\cite{Oga2014}, long carrier diffusion lengths\cite{Samuel2013,Guichuan2013,dong2015electron,shi2015low} and
strong absorption coefficients\cite{Wolf2014}. The power conversion efficiency (PCE) of HP solar cells has reached $29.5\%$\cite{nrel}.
However, the limitation of toxicity and instability still need to be overcome for further improvement.

Lead-free halide double perovskites (HDPs) with formula $A_2BB^{'}X_6$ have been proposed as environmentally
friendly alternatives \cite{Volonakis.2017, Luo.2018,zhou2019manipulation,Ning.2019, Majher.2019,Bartel.2020} to HPs with long working lifetimes and have been applied in
various optoelectronic devices, e.g. LEDs\cite{Luo.2018,Majher.2019}, photocatalysts\cite{Zhou2018} and solar cells\cite{Zhao.2017,Greul.2017,Gao2018,Pantaler2018}. For most HDPs, $A$ is chosen as a Cs$^+$ or organic
cation with a large ionic radius to stabilize the crystal structure, while $X$ represents a halide ion. $B$ and $B'$ are occupied by monovalent and trivalent elements, respectively, such as Ag$^+$, Na$^+$ and
Bi$^{3+}$, In$^{3+}$. Compared to conventional HPs, most HDPs have large band gaps which limit their applications in the visible region.
The band structures can be engineered by chemical substitution, alloying and doping
strategies\cite{Volonakis.2016, Luo.2018, Majher.2019, zhou2019manipulation, Bartel.2020}.
Density functional theory (DFT) is a powerful tool to study the geometric and electronic structures of the materials, and therefore
may provide useful guidance for the  band structures engineering of DHPs.

In this work, we investigate five HDPs, including Cs$_2$AgInCl$_6$,  Cs$_2$NaInCl$_6$, Cs$_2$AgBiCl$_6$, Cs$_2$AgBiBr$_6$ and Cs$_2$NaBiCl$_6$,  via DFT calculations. We compare the results obtained by the Perdew-Burke-Ernzerhof (PBE) semilocal functional~\cite{perdew1996} and Heyd-Scuseria-Ernzerhof (HSE) hybrid functional\cite{heyd2003hybrid, krukau2006influence}. The results show that HSE can predict much better geometries of the DHPs, including the lattice constants and the $B$-$X$ bond lengths. Importantly, we find that the electronic structures of DHPs are very sensitive to their geometries, especially the $B$-$X$ bond lengths. As a consequence, the electronic structures calculated by the HSE functional using the PBE optimized structures may still significantly underestimate the band gaps, whereas the calculations based on the HSE optimized geometries can give much better results. The sensitivity of the band gaps of the DHPs to their geometries opens a promising path for  band structure engineering by doping and alloying.

%\section{Computational Details}

The first-principles calculations are carried out using the Atomic-orbital Based Ab-initio Computation at USTC (ABACUS) package
\cite{chen2010systematically, li2016large, abacus}. We use both the PBE and HSE functionals  in the calculations.
In the HSE calculations, the mixing parameter of the Hartree-Fock exchange, $\alpha$=0.25 is used for DHPs with $B$=Na, and  $\alpha$=0.4 is used for  other DHPs,
to obtain good agreement of the band gaps with the experiments.
The spin-orbit coupling (SOC) effects are  taken into account to calculate the band structures.

We adopt the SG15 \cite{schlipf2015optimization} optimized norm conserving Vanderbilt-type (ONCVP) pseudopotentials\cite{Hamann2013}, where Cs: 5$s^2$5$p^6$6$s$,  Ag: 4$s^2$4$p^6$4$d^{10}$5$s$,  In: $4d^{10}5s^25p$, Cl: $3s^23p^5$, Bi $5d^{10}6s^26p^3$ and Br: $4s^24p^5$ electrons are treated as valence electrons.
The second generation NAO bases, namely the DPSI bases sets~\cite{lin2021strategy}, are used in all calculations. More specifically,
Cs with [4$s$ 2$p$ 1$d$], Ag with [4$s$ 2$p$ 2$d$ 1$f$], In with [2$s$ 2$p$ 2$d$], Cl with [2$s$ 2$p$ 1$d$], Na with [4$s$ 2$p$ 1$d$], Bi with [2$s$ 2$p$ 2$d$], Br with [2$s$ 2$p$ 1$d$] NAOs are used. The 8$\times$8$\times$8 $\Gamma$-centered Monkhorst-Pack $\boldsymbol{k}$-point mesh
is used for the self-consistent calculations, and a $\rm 12\times 12\times 12$  $\boldsymbol{k}$-point mesh is used to calculate the projected density of states (PDOS).

%\section{Results}

\begin{figure}
    \centering
    \includegraphics[width=0.5\textwidth]{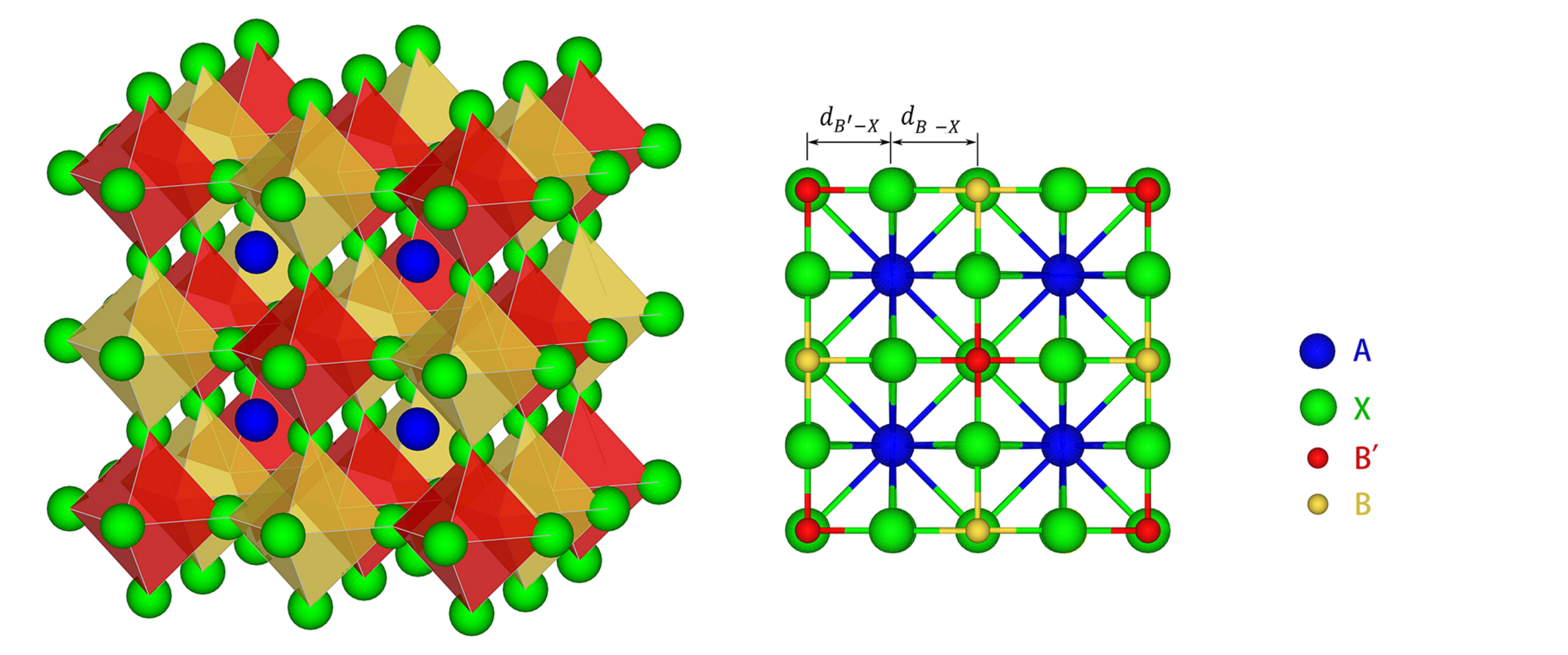}
    \caption{Crystal structure of $A_2BB'X_6$ with bond lengths $d_{B-X}$ and $d_{B'-X}$.}
    \label{fig:stru}
\end{figure}

Figure~\ref{fig:stru} depicts the  crystal structure of $A_2BB'X_6$ HDPs, which have $\rm Fm\overline{3}m$ space group symmetry.
The $B$ and $B' $ ions are located alternately in the center of the $A$$X_6$ octahedra. The $B$-$X$  ($B{'}$-$X$) bond length, i.e. the distance between $B$ ($B^{'}$) and $X$ in the same octahedral units is marked as $d_{B-X}$ ($d_{B'-X}$).

In previous first-principles calculations, the electronic structures of the HDPs calculated by hybrid functionals (HSE and PBE0) are usually performed
on the geometries obtained by the LDA/GGA functionals \cite{Savory.2016,Volonakis.2016,Zhao.2017,Zhou.2017}
partly because the computational cost of HSE is very expensive. It has been shown that the electronic structures of DHP can be sensitive to their geometries\cite{Volonakis.2017,zhou2019manipulation, Ning.2019}.
In this study, we would like to examine the effects of the HSE on the geometries of the DHPs,
and consequently the impact on their electronic structures.

We first calculate the lattice constants of the five HDPs by the PBE and HSE functionals, and the results are compared with the experimental values in Table~\ref{table:1}.
The lattice constants are fitted by the Birch-Murnaghan~\cite{birch1947finite} equation of state (EOS).
 The mean absolute errors (MAEs) of the lattice constants calculated from the PBE and HSE functionals, with respect to the experimental values, are 0.197 \AA ~and 0.120 \AA, respectively.
 As we see, the lattice constants obtained by HSE are in much better agreement with the experimental values than those by PBE.
 We also compare the equilibrium bond length $d^0_{B-X}$ calculated by the PBE and HSE functionals. In order to reduce the degrees of freedom in the comparisons and to compare with the experimental results, the lattice constants are fixed to the experimental values. We also fix the center of $B$ and adjacent $X$ atoms within the same octahedra, and then move the two atoms to change the distance between them and obtain a series of geometries. During the movement, the distance between $B$ and $B'$ is fixed, which means that the length increased by $d_{B-X}$ is equal to the length reduced by $d_{B'-X}$, thus, the symmetry is maintained. We estimate equilibrium bond lengths according to energy-versus-distance [$E(d)$] curves of five solids by using two functionals. An example of $E(d)$  can be found in the supplemental materials.  The results are listed in Table~\ref{table:2}, and compared with the experimental values.
 The PBE significantly underestimates the $B$-$X$ bond lengths, whereas
 HSE yields much better $B$-$X$ bond lengths with a MAE of 0.008 \AA, which is significantly less than the MAE of the PBE functional 0.038 \AA .

\begin{table}[t]
    \centering
    \caption{HSE lattice constants (in \AA) were fitted by Birch-Murnaghan \cite{birch1947finite} equation of state (EOS). }
    \label{table:1}
    \begin{tabular}{ c c c c }
        \hline\hline
            & PBE & HSE & Experiment \\
        \hline
        $\rm Cs_2AgInCl_6$ & 10.678 & 10.564 & 10.481\cite{Volonakis.2017}\\
        $\rm Cs_2NaInCl_6$ & 10.711 & 10.599 & 10.514\cite{Noculak.2020}, 10.534\cite{zhou2019manipulation}\\
        $\rm Cs_2AgBiCl_6$ & 10.974 & 10.946 & 10.777\cite{Volonakis.2016,McClure.2016}\\
        $\rm Cs_2AgBiBr_6$ & 11.447 & 11.380 & 11.250\cite{Slavney.2016}\\
        $\rm Cs_2NaBiCl_6$ & 11.036 & 10.966 & 10.839\cite{morrs1972crystal}, 10.842\cite{zhou2019manipulation}\\
        \hline
        MAE            & 0.197  & 0.120  &             \\
        \hline\hline
    \end{tabular}
\end{table}

\begin{table}[H]
    \centering
    \caption{Comparison of the equilibrium bond lengths $d_{B-X}^{0}$(in \AA) obtained
    by PBE and HSE functionals with the experimental values. }
    \label{table:2}
    \begin{tabular}{ c c c c }
        \hline\hline
            & PBE & HSE & Experiment \\
        \hline
        $\rm Cs_2AgInCl_6$ & 2.683 & 2.728 & 2.733\cite{Volonakis.2017}\\
        $\rm Cs_2NaInCl_6$ & 2.721  & 2.743 & 2.748 \cite{Noculak.2020}, 2.763\cite{zhou2019manipulation}\\
        $\rm Cs_2AgBiCl_6$ & 2.670  & 2.698 & 2.707\cite{Volonakis.2016,McClure.2016}\\
        $\rm Cs_2AgBiBr_6$ & 2.775 & 2.797 & 2.803\cite{Slavney.2016}\\
        $\rm Cs_2NaBiCl_6$ & 2.703 & 2.736 & 2.754\cite{morrs1972crystal, zhou2019manipulation}\\
        \hline
        MAE            & 0.038 & 0.008 &            \\
        \hline\hline
    \end{tabular}
\end{table}

\begin{figure}[htbp]
    \centering
    \includegraphics[width=0.5\textwidth]{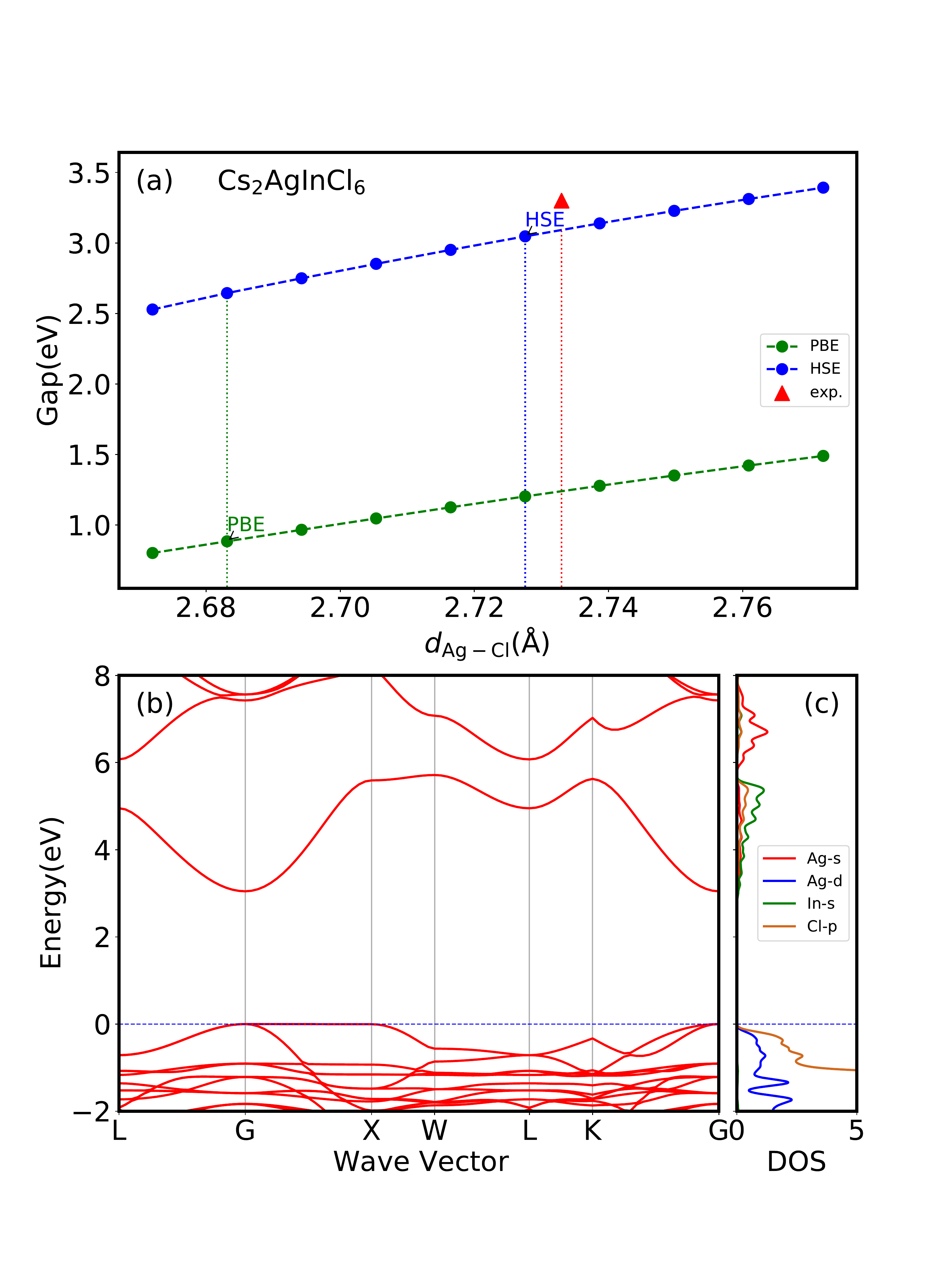}
    \caption{(a) Band gaps of $\rm Cs_2AgInCl_6$ with a lattice constant of 10.481 \AA ~predicted by PBE and HSE as a function of bond length $ d_{\rm Ag-Cl}$. The experimental structure and band gap were obtained from Ref. \citenum{Volonakis.2017}.
        (b)-(c) Band structure and partial DOS of $\rm Cs_2AgInCl_6$ with $ d_{\rm Ag-Cl}^{0, HSE}$ calculated using the HSE functional.}
        \label{fig:CsAgInCl}
\end{figure}

To see how the crystal structures affect the electronic  structures in the DHP materials, we calculate the band gaps as functions of
the $B$-$X$ bond lengths.

Figure \ref{fig:CsAgInCl}(a) depicts the band gaps as functions of Ag-Cl bond length calculated by PBE and HSE functionals for Cs$_2$AgInCl$_6$.
The experimental lattice constant is used.
Cs$_2$AgInCl$_6$ shows a negligible SOC effect
in our calculations,
and therefore, we show only the results without SOC.
As we see from the figure, the HSE calculated band gap increases notably with increasing of Ag-Cl bond length, $d_{\rm Ag-Cl}$.
Even though the PBE functional significantly underestimates the gap, it predicts the similar trends of the band gaps changing with  Ag-Cl bond length.
The band gap calculated by HSE at the PBE  Ag-Cl bond length (2.683 \AA) is approximately 2.644 eV, which is significantly lower than the experimental value (3.3 eV),
marked by the red triangle in the figure.
In contrast, the band gap calculated at the HSE optimized Ag-Cl bond length (2.728 \AA) is approximately 3.047 eV,
which is in excellent agreement with the experimental value.
These results suggest that the band gap calculated by HSE on the geometry obtained by the PBE
functional would still significantly underestimate the band gap.

Figure \ref{fig:CsAgInCl}(b) shows the band structure of Cs$_2$AgInCl$_6$  calculated  by the HSE functional,
using the  $d_{\rm Ag-Cl}$ also obtained by the HSE.
The corresponding PDOS are shown in Fig.~\ref{fig:CsAgInCl}(c).
The lowest conduction band (CB) of Cs$_2$AgInCl$_6$ is very dispersive, since it  is dominated by the delocalized Cl 3$p$ and In 5$s$ electrons
with some small contribution from the Ag 5$s$ states.
The dispersive bands suggest that the electrons have very small effective masses, and therefore
large carrier mobility, which is good for photoelectric device  applications.
The highest VB is much flatter, especially there is a flat band between the $\Gamma$ and $X$ points. This flat band originates from the hybridization of Cl 3$p$,
and Ag 4$d$ electrons [Fig. \ref{fig:CsAgInCl}(c)] near the Fermi level .

We also calculate the band gaps as a function of  Na-Cl bond length for Cs$_2$NaInCl$_6$ using standard HSE with mixing parameter of 0.25, and find that
the trend of band gaps over the Na-Cl bond length is very similar to that of Cs$_2$AgInCl$_6$.
More results of the electronic structures for Cs$_2$NaInCl$_6$ can be found in the supplemental materials.

\begin{figure}[tbp]
    \centering
      \includegraphics[width=0.5\textwidth]{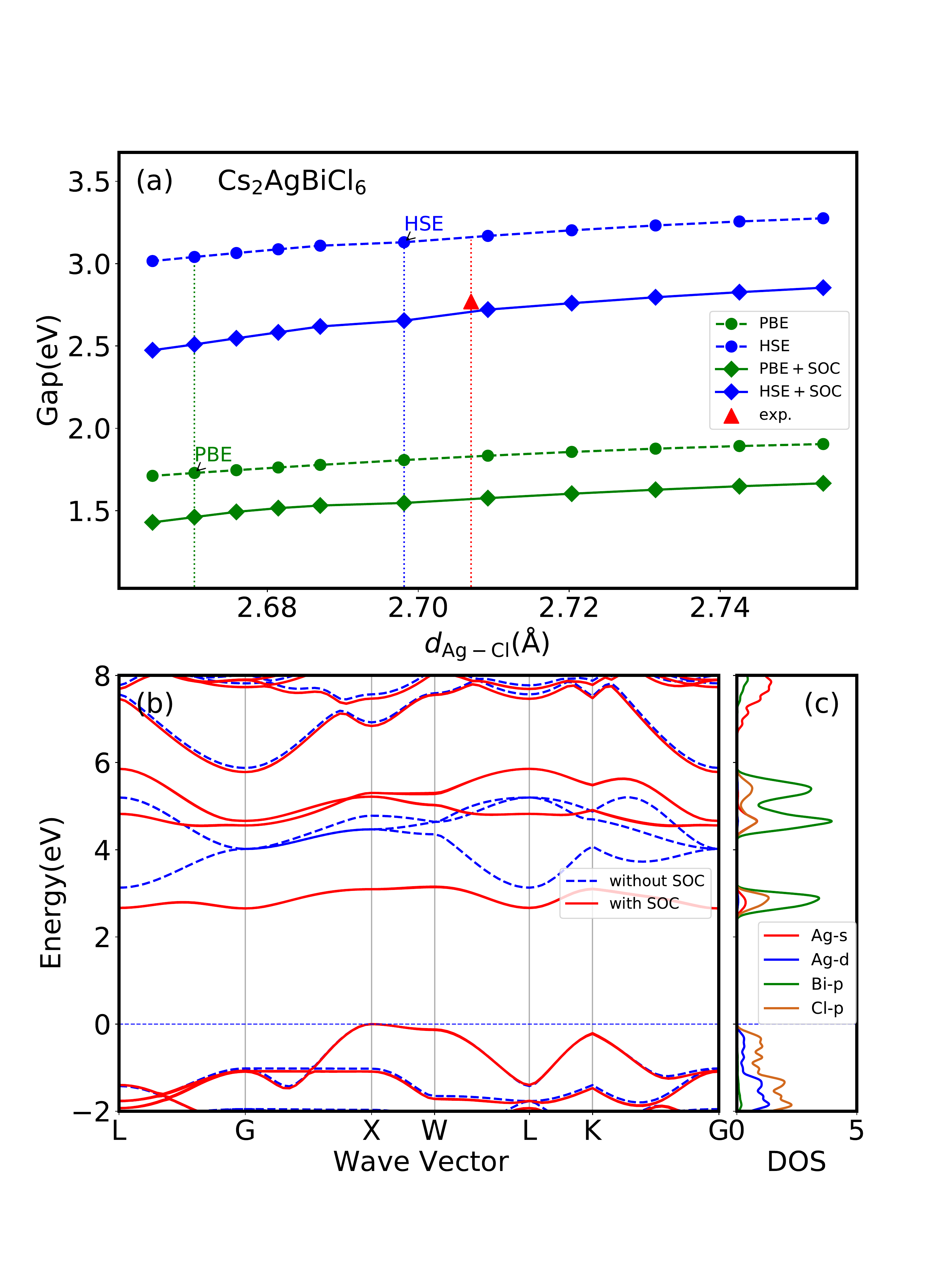}
    \caption{(a) Band gaps of Cs$_2$AgBiCl$_6$ with a lattice constant of 10.777 \AA ~predicted by four methods(i.e. PBE, PBE+SOC, HSE and HSE+SOC) as a function of bond length $d_{\rm Ag-Cl}$.
        The experimental structure and band gap were obtained from Ref. \citenum{McClure.2016}.
        (b)-(c) Band structure and partial DOS with SOC of $\rm Cs_2AgBiCl_6$ with $d_{\rm Ag-Cl}^{0, HSE}$ calculated using the HSE+SOC functionals.}
        \label{fig:Cs2AgBiCl6}
\end{figure}

 Substituting In with Bi in Cs$_2$AgInCl$_6$ would be expected to result in very different band structures.
 $\rm Cs_2AgBiCl_6$ has a much larger lattice constant of 10.777 \AA ~than $\rm Cs_2AgInCl_6$,
 and the Bi ion introduces strong SOC effects.
Indeed, Cs$_2$AgBiCl$_6$ has an indirect band gap, in contrast to Cs$_2$AgInCl$_6$.
The feature of the indirect band gap of this compound is captured by both PBE and HSE functionals, with or without SOC.
We plot the (indirect) band gaps as functions of Ag-Cl bond length for both PBE and HSE functionals.
Similar to the case of Cs$_2$AgInCl$_6$, both PBE and HSE functionals predict that the band gap increases with increasing
$d_{\rm Ag-Cl}$ for Cs$_2$AgBiCl$_6$, although the magnitude is less dramatic, as shown in Fig.~\ref{fig:Cs2AgBiCl6}(a).
The PBE calculated band gap at the HSE calculated Ag-Cl bond length, is 1.807 eV without SOC, and decreases to 1.547 eV after turning on SOC, both
significantly underestimate the experimental band gap of 2.77 eV.
The band gap calculated by the HSE functional with nonstandard mixing parameter of 0.4 at this bond length is
approximately 3.130 eV without SOC, which
significantly overestimates the experimental band gap.
After turning on the SOC, the band gap decreases by approximately 0.5 eV, and is in excellent agreement
with the experimental gap marked as the red triangle in the figure.
However, if we calculate the band gap using the PBE optimized Ag-Cl bond length, the band gap would be 0.145 eV smaller.

The band structures of Cs$_2$AgBiCl$_6$ are shown in   Fig.~\ref{fig:Cs2AgBiCl6}(b). The band structures are calculated by HSE, with (red solid lines)
and without (blue dashed lines) SOC. The experimental lattice constant is used, and the Ag-Cl bond length is determined from the HSE calculations.
As shown in the figure, the  strong SOC of the Bi atoms leads to giant splitting of the conduction bands, which significantly
reduces the band gap. The direct band gap near the X point is approximately 3.092 eV, which is also in good agreement
with the experimental value of 3.3 eV\cite{Volonakis.2016}.

Notably, the lowest CBs become much less dispersive compared to those without turning on SOC.
Especially, the lowest CB becomes very flat after turning on SOC.
This is different from the Pb-based perovskites, where SOC actually reduces the electron effective masses.
PDOS [see Fig.~\ref{fig:Cs2AgBiCl6}(c)] analysis shows that the lowest conduction bands are
made up mainly of Bi-6$p$ states hybridized with Ag-5$s$ and Cl-3$p$ states,
which split into two distinct peaks in PDOS under SOC.
Shi and Du suggested that the flat CB in the Bismuth DHPs is due to the large
electronegativity difference among cations and the large nearest-neighbor distances in cation sublattices~\cite{Shi.2015}.
Savory et al. argued that the flat CB band is due to a mismatch in angular momentum of the frontier atomic orbitals of the $B$ and $B'$ ions
\cite{Savory.2016}.
Here we show that the SOC also plays a crucial role for the flat CB bands in the Bismuth DHPs.

The band structrues of Cs$_2$NaBiCl$_6$ and Cs$_2$AgBiBr$_6$  have very similar features to those of
Cs$_2$AgBiCl$_6$. Detailed results for the two materials are given in the Supplementary Materials.

%\section{Summary}

We investigate the geometries and electronic structures of Cs$_2$$B$$B'$$X_6$  DHPs via first-principles calculations. We compare the results
given by the PBE and HSE functionals. The results show that HSE can predict much better geometries of the DHPs, including the lattice constants and the $B$-$X$ bond lengths. Importantly, the electronic structures of DHPs are very sensitive to their geometries, especially the $B$-$X$ bond lengths. Therefore the electronic structures calculated by the HSE functional using the PBE optimized structures may still significantly underestimate the band gaps, whereas the calculations on the HSE optimized geometries can give much better results. The results indicate that mitigating the strong self-interaction errors are crucial for correctly describing not only the electronic structure, but also the geometries of DHP materials. The sensitivity of  band gaps of the DHPs to their geometries opens a promising path for band structure engineering by doping and alloying.

\begin{acknowledgement}
This work was funded by the Chinese National Science
Foundation Grant Numbers 12134012, 11874335, and 12188101. The numerical calculations were performed on the USTC HPC facilities.
\end{acknowledgement}

\begin{suppinfo}

%A listing of the contents of each file supplied as Supporting Information
%should be included. For instructions on what should be included in the
%Supporting Information as well as how to prepare this material for
%publications, refer to the journal's Instructions for Authors.

The following files are available free of charge.
\begin{itemize}
  \item sm.pdf: Supplementary results for the paper.
\end{itemize}

\end{suppinfo}

%\bibliographystyle{apsrev4-2}%
%\bibliography{double-hp}
\providecommand{\latin}[1]{#1}
\makeatletter
\providecommand{\doi}
  {\begingroup\let\do\@makeother\dospecials
  \catcode`\{=1 \catcode`\}=2 \doi@aux}
\providecommand{\doi@aux}[1]{\endgroup\texttt{#1}}
\makeatother
\providecommand*\mcitethebibliography{\thebibliography}
\csname @ifundefined\endcsname{endmcitethebibliography}
  {\let\endmcitethebibliography\endthebibliography}{}

\end{document}